\documentclass{PoS}
\usepackage[dvips]{epsfig}
\def\lsim{\raise0.3ex\hbox{$<$\kern-0.75em\raise-1.1ex\hbox{$\sim$}}}
\def\gsim{\raise0.3ex\hbox{$>$\kern-0.75em\raise-1.1ex\hbox{$\sim$}}}

\title{Recent lattice results on finite temperature and density QCD, part I}

\ShortTitle{QCD thermodynamics I}

\author{\speaker{Frithjof Karsch}\thanks{This work has been supported in 
part by contract DE-AC02-98CH10886 with the U.S. Department of Energy.}\\
  Physics Department, Brookhaven National Laboratory, Upton, NY 11973, USA\\
      E-mail: \email{karsch@bnl.gov}}

\abstract{
We discuss recent progress made studies of bulk thermodynamics of 
strongly interacting matter through lattice simulations of QCD with an 
almost physical light and strange quark mass spectrum. We present
results on the QCD equation of state at vanishing and non-vanishing 
quark chemical potential and show first results on baryon number and 
strangeness fluctuations, which might be measured in event-by-event 
fluctuations in low energy runs at RHIC as well as at FAIR.
}

\FullConference{Critical Point and Onset of Deconfinement
          4th International Workshop\\
		 July 9-13 2007\\
		 GSI Darmstadt,Germany}

\begin{document}

\section{Introduction}

The striking observations of large elliptic flow and jet quenching
at RHIC \cite{RHIC} hints at an early equilibration of the dense 
matter formed in relativistic heavy ion collisions. This in turn
suggests that the constituents of this medium, which at RHIC equilibrates 
at temperatures less than about $300$~MeV or energy densities of 
about $15$~GeV/fm$^3$, are still interacting strongly. The 
equation of state (EoS) describing this medium thus should exhibit significant
deviations from ideal gas behavior. This 
has indeed been seen in calculations of the QCD equation of state on
the lattice, where deviations from the non-interacting limiting behavior,
$\epsilon=3p$, show up as a large peak in the trace anomaly in units of $T^4$, 
$\Theta^{\mu\mu}(T)/T^4\equiv (\epsilon -3p)/T^4$. 

Pretty soon bulk thermodynamics, and in particular the  EoS, 
will be probed at even higher temperatures at the LHC. This will be 
complemented by experimental studies of matter at lower temperature but 
non-zero baryon number density through low energy heavy ion collisions at RHIC 
and at the future European heavy ion facility, FAIR, at GSI/Darmstadt. It thus
will be of importance to get control over the EoS in a wide range 
of temperatures and to analyze also its dependence on the baryon number 
density or the baryon chemical potential.

While good control over the continuum extrapolation of numerical results 
on the EoS has been reached in lattice
calculations for the SU(3) gauge theory, much work is still needed to
get similarly good control over calculations performed with physical light
and strange quark masses. We will discuss here the current status of these
lattice calculations.

During recent years the steady progress made in lattice calculations 
through important new algorithmic developments \cite{rhmc} and increased 
computing resources helped to greatly advance our knowledge of the 
thermodynamics of strongly interacting elementary particles \cite{Schladming}. 
On the one hand, lattice calculations of the equation of state
\cite{aoki,milc_eos,p4_eos} and transition temperature 
\cite{milc_Tc,p4_Tc,aoki_Tc} with an almost realistic quark 
mass spectrum are now possible at vanishing chemical potential.
On the other hand, lattice calculations based on various approximation
schemes now also provide first insight into the thermodynamics at 
non-zero quark or baryon chemical potential ($\mu_q$) \cite{muref}. 

In this review we will discuss some of the results obtained recently
with ${\cal O}(a^2)$ improved staggered fermion formulations. As these
results have been presented in July/August of 2007 in a very similar 
format at the '4th International Workshop on Critical Point and Onset of 
Deconfinement' and at the 'XXV International Symposium on Lattice Field
Theory' the write-up of these talks has been splitted into two parts.
In this first part we will focus on recent results on the equation of 
state at vanishing
and non-vanishing chemical potential and discuss results on strangeness
and baryon number fluctuations which may become detectable in heavy
ion collisions. Part II \cite{partII} focuses on a discussion of results 
on the nature of the QCD transition and reports on the current status of
calculations of the transition temperature in QCD.

\section{The QCD equation of state}

While continuum extrapolated results for the EoS of the 
SU(3) gauge theory have been established more than 10 years 
ago \cite{Schladming}, this still has to be achieved for QCD with a 
physical quark mass spectrum.
Lattice calculations with light dynamical quarks, of course, introduce their 
own set of problems into the numerical study of QCD thermodynamics. 
The well-known problems with the discretization of fermion actions that
require to break the chiral symmetry of the continuum action and/or requires
the introduction of additional fermion species leads to a cut-off 
dependent modification of the hadron spectrum that will influence thermodynamic
properties at low temperatures, {\it i.e.} in the hadronic phase.
This introduces additional cut-off dependencies
into lattice studies of QCD thermodynamics that are not present in pure gauge 
theories and require a detailed analysis. Aside from this, however, studies
of QCD thermodynamics suffer from similar discretization errors as pure gauge 
theories which strongly influence studies of bulk thermodynamics in the plasma
phase of QCD and, in particular, show up as a strong cut-off dependence of the
high temperature limit of bulk thermodynamic observables. These cut-off
effects essentially arise from the need to replace derivatives in the QCD
Lagrangian by finite
differences. In the next subsection we briefly recall the importance of 
these effects for the analysis of thermodynamics in the high temperature 
phase of SU(3) gauge theories before starting to discuss the thermodynamics 
of QCD with two degenerate, light quark masses and a heavier strange quark 
mass.

\subsection{Cut-off effects in QCD thermodynamics}

\subsubsection{SU(3) gauge theories revisited}
 
\begin{figure}[t]
\begin{center}
\hspace*{-0.4cm}\epsfig{file=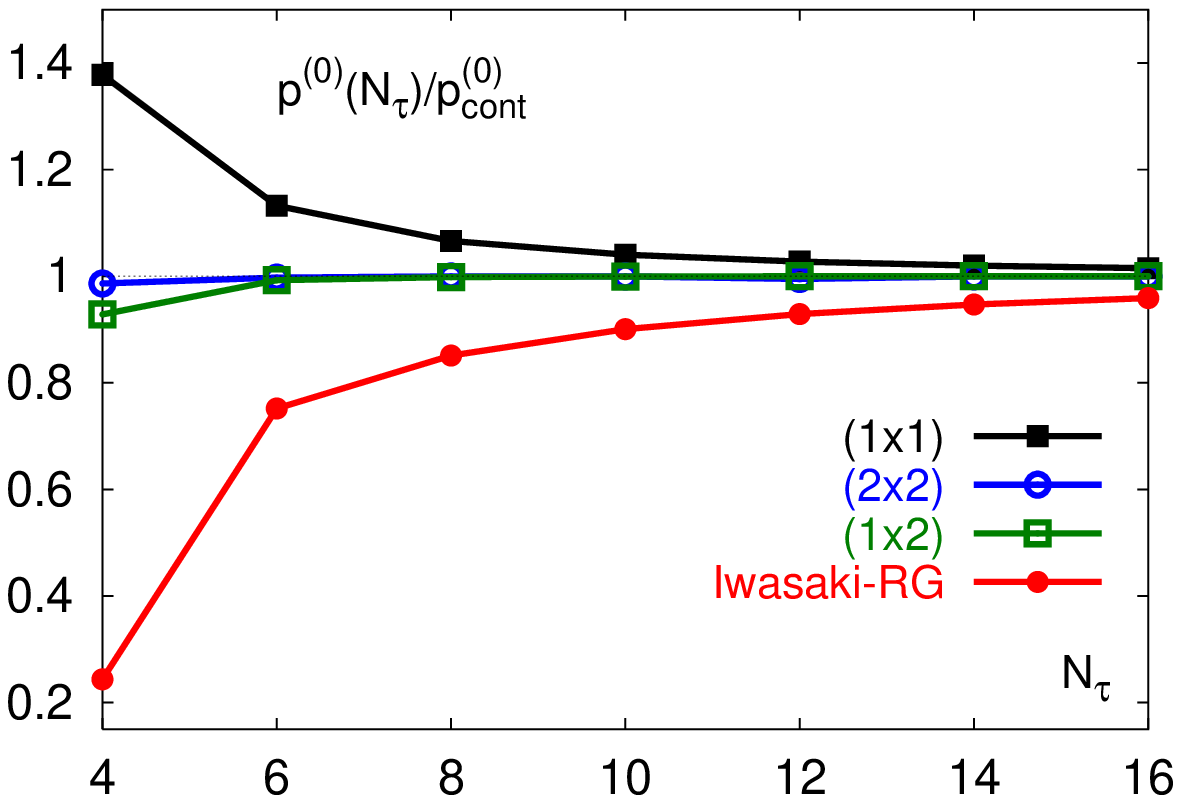, width=7.9cm}
\hspace*{-0.4cm}\epsfig{file=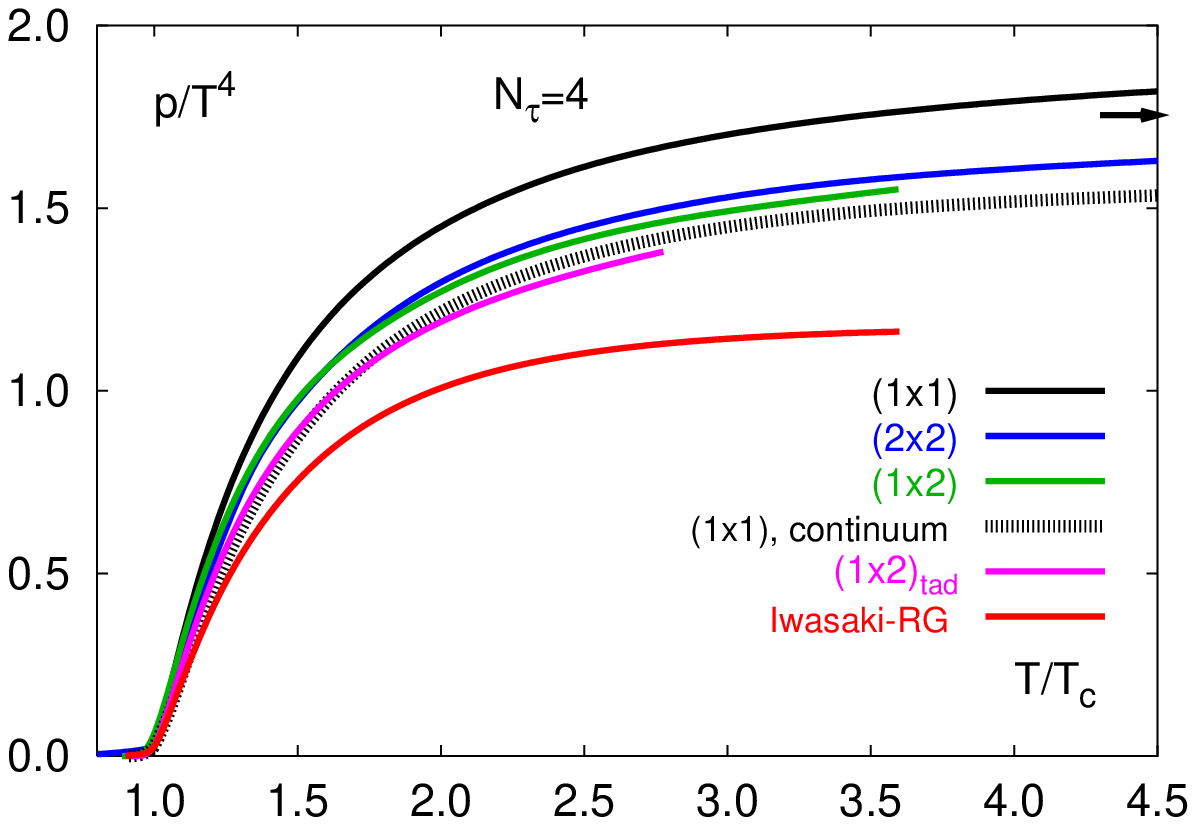,width=7.9cm}
\end{center}
\caption{Pressure in the SU(3) gauge theory calculated for the standard
Wilson action ($1\times 1$) and several improved actions analytically
in the infinite temperature, ideal gas limit on lattices with temporal
extent $N_\tau$ (left). The right hand figure shows numerical results
obtained on lattices with temporal extent $N_\tau=4$. 
}
\label{fig:SU3_p}
\end{figure}

In the standard approach to the discretization of a SU(3) gauge theory the
Euclidean action is represented by a sum over ($1\times 1$)-Wilson loops 
(plaquettes). This introduces ${\cal O}(a^2)$ discretization errors. It is
well known how to eliminate these leading order discretization errors on the
tree level by adding larger Wilson loops, eg. ($1\times 2$) or ($2\times 2$)
loops. This shifts the leading order discretization errors to ${\cal O}(a^4)$.
Other improvement schemes, like for instance the Iwasaki action, improve
certain non-perturbative features of the lattice action, but do not aim at
an elimination of the  ${\cal O}(a^2)$ errors. These discretization errors
also show up in lattice studies of thermodynamic quantities on lattices
with finite temporal extent ($N_\tau$); as the relevant scale is the lattice 
cut-off in units of the temperature, 
$aT\equiv 1/N_\tau$, ${\cal O}(a^n)$ cut-off errors thus show up in
finite temperature calculations as ${\cal O}(1/N_\tau^n)$ errors.
This can explicitly been analyzed in the infinite temperature ideal gas limit
of QCD. In Fig.\ref{fig:SU3_p}(left) we show the cut-off dependence
of the pressure calculated analytically for various SU(3) actions as 
function of $N_\tau$ and normalized to the continuum 
($N_\tau\rightarrow \infty$) Stefan-Boltzmann value.   

One lesson learned in studies of the EoS of a pure gauge theory is that it
is extremely important to use at least ${\cal O}(a^2)$ improved gauge actions 
for the calculation of bulk thermodynamic observables like eg. energy density 
and pressure, if one wants to reduce cut-off effects  at high 
temperature already on lattices with small temporal extent. 
As can be seen in Fig.~\ref{fig:SU3_p}(right) also at temperature close 
to the deconfinement transition temperature cut-off effects on lattices 
with fixed temporal extent ($N_\tau=4$) follow the pattern seen already
in the infinite temperature limit. While the pressure calculated with 
actions that have ${\cal O}(a^2)$ cut-off errors deviates strongly from 
the continuum extrapolated result also in the
vicinity of the transition temperature calculations with ${\cal O}(a^2)$ 
improved gauge actions yield results on coarse latices that are already 
close to the continuum extrapolated result. 
 
\begin{figure}[t]
\begin{center}
\hspace*{-0.4cm}\epsfig{file=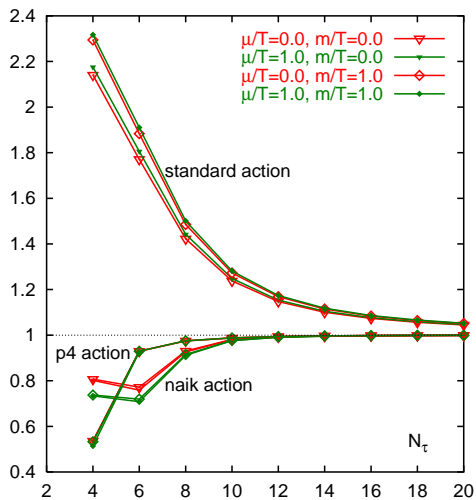,width=6.5cm}
\end{center}
\caption{Cut-off dependence of the infinite temperature ideal gas pressure
in staggered fermion formulations of QCD thermodynamics. Shown are 
analytic results for the pressure calculated in the thermodynamic limit
($N_\sigma\rightarrow \infty$) on lattices with temporal extent $N_\tau$ 
for the standard 1-link action as well as two ${\cal O}(a^2)$ improved
actions. Results are shown for vanishing and non-vanishing values
of the chemical potential ($\mu/T = 0,\; 1$).
}
\label{fig:pfree}
\end{figure}

\subsubsection{Thermodynamics with staggered fermions}

These generic features also hold true in calculations performed in QCD with 
light quarks. The standard discretization scheme of the fermionic part of 
the QCD action (1-link actions) introduces ${\cal O}(a^2)$ errors that can
be removed in thermodynamic observables on the tree level by introducing 
suitably chosen additional 3-link terms (Naik, p4). The resulting
cut-off effects in the infinite temperature, ideal gas limit are shown
in Fig.~\ref{fig:pfree}. We note that the improvement schemes improve
bulk thermodynamic observables at vanishing as well as non-vanishing
chemical potential. These actions are usually supplemented
with additional terms that reduce the flavor symmetry breaking (smearing)
and/or reduce cut-off effects beyond the tree level (tadpole 
improvement)\footnote{For details on staggered fermion actions used in 
large scale studies of QCD thermodynamics we refer to the literature on, 
eg.  the 1-link, stout smeared action \cite{aoki_Tc}, the asqtad action 
\cite{milc_Tc} and the p4fat3 action \cite{p4_Tc} (and references therein).}.
None of these improvements, however, alter the behavior in
the infinite temperature limit. They are thus not expected to be essential
for studies of bulk thermodynamics in the high temperature phase of QCD. 

The early calculations with unimproved staggered fermions \cite{milc_ss} and 
calculations with unimproved Wilson fermions \cite{Wilson} as well as 
recent calculations with the 1-link, stout smeared action action \cite{aoki}
indeed showed discretization errors similar to those observed in pure gauge 
theories; 
at high temperature energy density and pressure deviate strongly 
from the continuum ideal gas behavior and reflect the large cut-off dependence
of the ideal quark-gluon gas in these discretization 
schemes\footnote{It sometimes is attempted to correct for these cut-off effects
by normalizing numerical results obtained at some value of the temperature
in simulations on lattices with
given temporal extent $N_\tau$ to the Stefan-Boltzmann value
for pressure and energy density on the same size lattices at infinite 
temperature. This, however,
does not solve the problem as cut-off effects are known to be temperature
dependent. Moreover, it leads to thermodynamic inconsistencies when calculating,
for instance, the trace anomaly, $(\epsilon -3p)/T^4$, which vanishes in the
infinite temperature, ideal gas limit.}. We therefore will focus in the 
following on a discussion of results on bulk thermodynamics obtained with 
${\cal O}(a^2)$ improved gauge and fermion actions.

\subsection{QCD Equation of State at vanishing chemical potential}

\subsubsection{Bulk thermodynamics: Trace anomaly and entropy density}

Lattice calculations of the QCD equation of state with almost physical light
quark masses and a physical value of the strange quark mass have, so far,
been performed only on lattices with temporal extent $N_\tau =4$ and $6$.
These calculations have been performed on lines of constant physics (LCP), 
{\it i.e.} with bare quark masses chosen such that a set of hadron masses 
stays constant as the continuum limit is approached. In order to study QCD 
thermodynamics over a wide temperature range it is necessary to determine 
the parameters that characterize the LCP at zero temperature over a wide
range of lattice cut-offs. On the LCP one, furthermore, has to calculate
an additional observable that can be used to set the temperature scale
for the finite-T calculations. As this scale has to be determined at zero
temperature, {\it i.e.} on large lattices, for quite a
few different cut-off values $(a)$ corresponding to the temperature values
($T=1/N_\tau a $) one wants to analyze, it is advantageous to use a gluonic
observable, e.g. the static quark potential, that is easy to calculate, 
is not influenced strongly by the chiral sector of QCD and has a weak 
cut-off dependence. In this way it becomes possible to minimize
the influence of spurious zero temperature cut-off effects on thermodynamic
studies and to control the genuine cut-off and quark mass dependence of 
numerical calculations at finite temperature. 

\begin{figure}[t]
\begin{center}
\hspace*{-0.4cm}\epsfig{file=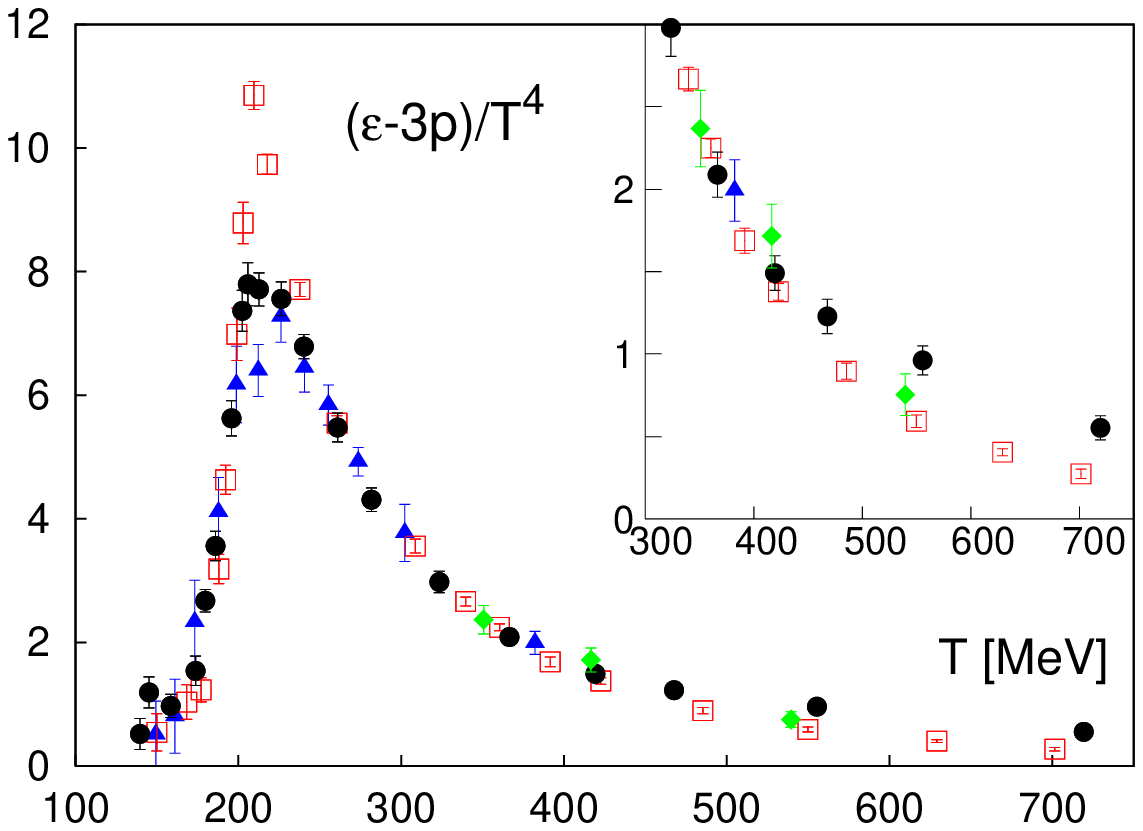, width=8.2cm}
\hspace*{-1.0cm}\epsfig{file=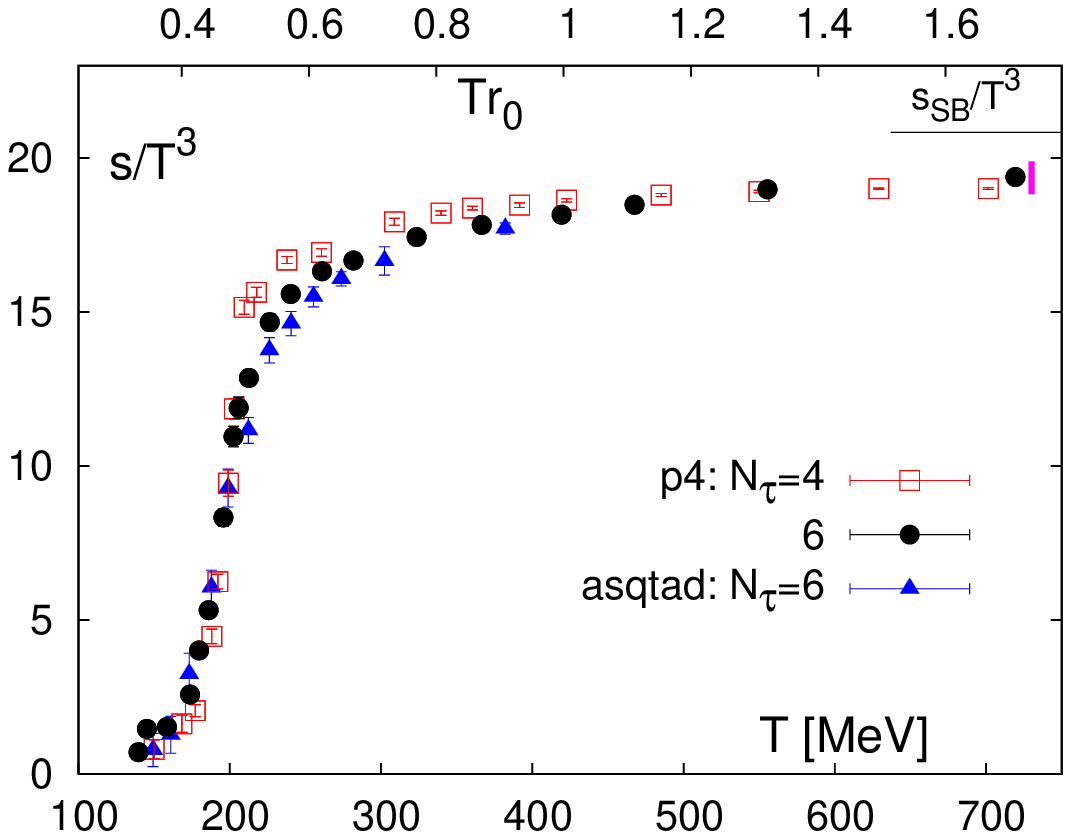,width=8.2cm}
\end{center}
\caption{The trace anomaly, $\Theta^{\mu\mu}(T)\equiv \epsilon -3p$, in units of $T^4$ 
(left) and the entropy density, $s\equiv \epsilon +p$, in units of $T^3$
calculated with the p4fat3 action \protect\cite{p4_eos} on lattices with 
temporal extent $N_\tau=4$ and $6$.
For $\Theta^{\mu\mu}/T^4$ we also show results on $N_\tau=8$ lattices 
(diamonds) obtained at high temperature. For $N_\tau=6$ results from 
calculations
with the asqtad action are also shown \protect\cite{milc_eos}. In the
right hand figure we also show the temperature scale $Tr_0$ (upper x-axis) 
which has been obtained from an analysis of static quark potentials at 
zero temperature \protect\cite{p4_eos}. The MeV-scale shown on the lower 
x-axis has been extracted from this using $r_0=0.469$~fm.
}
\label{fig:eps}
\end{figure}

In Fig.~\ref{fig:eps} we show results for the trace anomaly,
$\Theta^{\mu\mu}(T)/T^4 = (\epsilon -3p)/T^4$, and the entropy density,
$s/T^3 =(\epsilon + p)/T^3$, calculated with ${\cal O}(a^2)$ improved gauge 
and staggered fermion actions \cite{milc_eos,p4_eos}. We note that the entropy 
density
as well as energy density and pressure are not independent observables but 
are obtained from
$\Theta^{\mu\mu}(T)/T^4$ using standard thermodynamic relations, {\it i.e.}
after having evaluated the trace anomaly on the lattice, one obtains the
pressure through integration over the trace anomaly,
\begin{equation}
\frac{p}{T^4} - \frac{p}{T_0^4} = \int_{T_0}^{T} {\rm d}T' \frac{\epsilon -3p}{T'^5}  \; .
\label{pressure}
\end{equation}
Choosing the initial temperature $T_0$ deep enough in the hadronic phase 
where the pressure becomes exponentially small such that
the contribution of $p(T_0)/T_0^4$ can be ignored safely, one obtains the 
energy and entropy densities through suitable combinations of 
$\Theta^{\mu\mu}(T)$ and $p(T)$. The small vertical bar in the right hand
part of Fig.~\ref{fig:eps}(right) shows an estimate of the uncertainty
that arises from setting the pressure to zero at $T_0=100$~MeV. It is based
on an estimate of the entropy density in a hadron resonance gas at this 
temperature \cite{p4_eos}.

In Fig.~\ref{fig:eps}(left) we show results for the trace anomaly obtained in 
calculations with the p4fat3 action on lattices with temporal extent 
$N_\tau =4$, $6$ and $8$ on large spatial volumes, 
{\it i.e.} $V^{1/3}T = N_\sigma/N_\tau =4$ or larger.
The insertion in this figure shows the high temperature part of 
$(\epsilon -3p)/T^4$. Obviously cut-off effects are an issue in this regime.
For $T\gsim 400$~MeV, {\it i.e.} at temperatures larger than
about twice the transition temperature, results on the $N_\tau=4$ lattice
drop more rapidly than on the $N_\tau=6$ lattice. The calculations on the
$N_\tau =8$ lattice, however, confirm the latter results and suggest that
cut-off effects are under control in this high temperature region already
on the $N_\tau =6$ lattices. This is consistent with the analysis of 
cut-off effects for the p4-action in the infinite temperature, ideal gas
limit \cite{Hellerp4}. 

Also at $T \simeq 200$~MeV, {\it i.e.} at temperatures just above the 
transition region, the trace
anomaly shows some cut-off dependence. The peak in $(\epsilon -3p)/T^4$ is
much more pronounced in the $N_\tau=4$ calculations than in the $N_\tau=6$
case. This can be traced back to the rapid change of non-perturbative 
$\beta$-functions in the crossover region from weak to strong coupling 
\cite{p4_eos}. As can be seen from the entropy density shown in 
Fig.~\ref{fig:eps}(right) this leads to a noticeable cut-off dependence
in $s/T^3$ in this temperature range. It, however, has little effect 
on the high temperature behavior. This also holds true for energy
density and pressure \cite{p4_eos}. It also is reassuring that  
calculations on performed $N_\tau=6$ lattices with the p4fat3 action 
on quite large lattices, $V^{1/3}T=4$, are
in good agreement with simulations that have been performed with the asqtad 
action on a smaller physical volume, $V^{1/3}T =2$ \cite{milc_eos}. 
Results from the latter calculation are also shown in Fig.~\ref{fig:eps}.
The asqtad action has quite a different 
cut-off dependence at high temperature, it uses non-perturbatively improved
(tadpole) couplings and also incorporates a more sophisticated smearing
of 1-link terms in the staggered action to reduce flavor symmetry breaking
effects. The good agreement between asqtad and p4fat3 simulations thus
suggests that these features only play a minor role in the common temperature
range  explored in both calculations, 
$150 {\rm MeV} \lsim T \lsim 400 {\rm MeV}$.

The results shown in Fig.~\ref{fig:eps} have been obtained in calculations
with a physical strange quark mass and light quark masses that are about 
(2-2.5) times larger than in nature. This difference is of no significance
at high temperature as the quark masses are small in units of the 
temperature\footnote{The renormalization group invariant light quark mass
for the calculations performed with the p4fat3 action has been estimated to
be $m^{RGI}= 8.0(4)$~MeV \protect\cite{p4_eos}.}. It may, however, play a 
role in the low temperature hadronic phase. From the experience gained in 
simulations with different light quark masses \cite{milc_Tc,p4_Tc} it is 
to expected that the region of sudden rise in 
the trace anomaly as well as the entropy density shifts to somewhat smaller
temperatures in the case of physical quark mass values. Cut-off effects will
lead to a similar effect. This deserves a further careful analysis (see also
disccusion in part II \cite{partII}).

\subsubsection{Equation of state and velocity of sound}

\begin{figure}[t]
\begin{center}
\epsfig{file=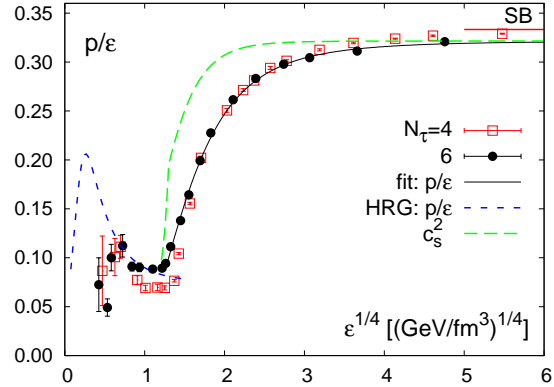,width=7.9cm}
\end{center}
\caption{The ratio $p/\epsilon$ as function of the fourth
root of the energy density obtained from calculations with the p4fat3
action on lattices with 
temporal extent $N_\tau =4$ and $6$. Also shown is the velocity of
sound extracted from a fit to $p/\epsilon$ \protect\cite{p4_eos} 
and using Eq.~\protect\ref{sound}.The dashed curve at low energies
shows the results for $p/\epsilon$ calculated in a hadron resonance gas
model (HRG).}
\label{fig:p_over_e}
\end{figure}

For the description of the expansion of dense matter created in heavy ion 
collisions, in particular its hydrodynamic modeling, the temperature 
dependence of bulk thermodynamic observables is not of direct interest.
It is more relevant to get good control over the dependence of the pressure 
on the energy density, $p(\epsilon)$, and deduce from this the velocity of 
sound,
\begin{equation}
c_s^2 = \frac{{\rm d} p}{{\rm d}\epsilon} = \epsilon 
\frac{{\rm d} p/\epsilon}{{\rm d}\epsilon} + \frac{p}{\epsilon}\; .
\label{sound}
\end{equation}
This is shown in Fig.~\ref{fig:p_over_e}. We note that the velocity of sound
gets close to the ideal gas value at energy densities of about $15$~GeV/fm$^3$
or temperatures $T\simeq 300$~MeV. However, for smaller energy densities, 
{\it i.e.} in the entire energy range of interest to the current RHIC 
experiments equation of state,  $p(\epsilon)$, as well as the velocity of 
sound drop rapidly. In the transition region,
{\it i.e.} at $\epsilon \simeq 1$~GeV/fm$^3$, one finds 
$c_s^2\simeq p/\epsilon \simeq 0.09$. 

We thus conclude that in the entire density regime, relevant
for the expansion of dense matter created at RHIC, the QCD EoS shows large
deviations from the conformal limit, $c_s^2 = p/\epsilon = 1/3$.

\subsection{Non-vanishing chemical potential}

\begin{figure}[t]
\begin{center}
\hspace*{-0.4cm}\epsfig{file=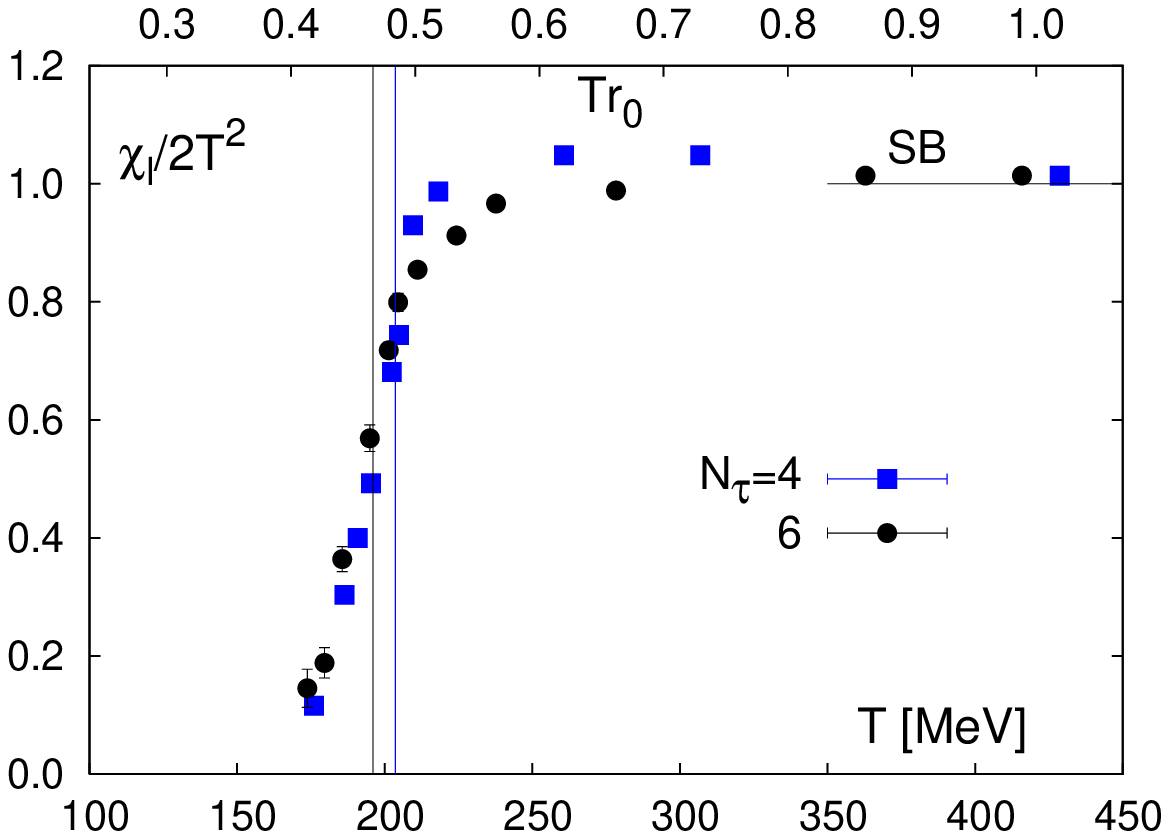, width=7.9cm}
\hspace*{-0.4cm}\epsfig{file=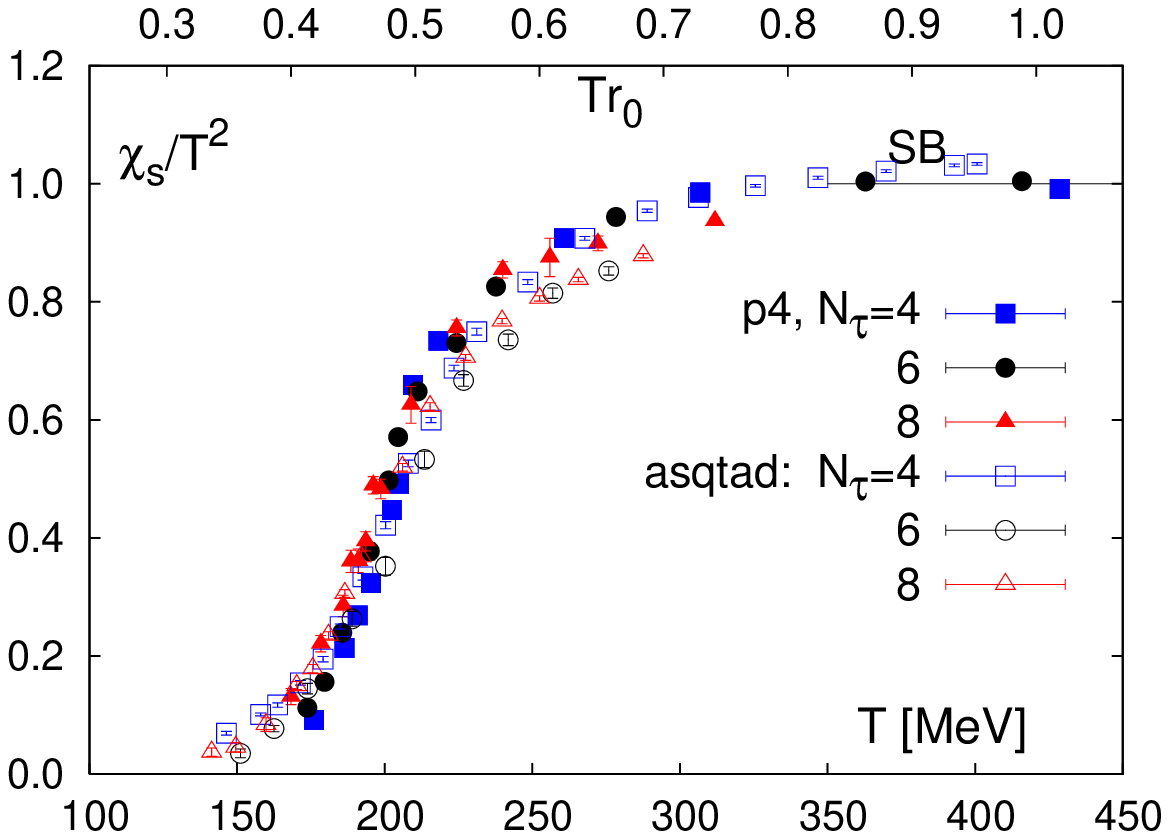,width=7.9cm}
\end{center}
\caption{Light ($\chi_l$) and strange ($\chi_s$) quark number susceptibilities
at vanishing chemical potential:
The left hand side shows $\chi_l/2T^2$ calculated with the p4fat3 action on 
lattices with temporal extent $N_\tau =4$ and $6$. Vertical lines indicate
here the location of transition temperatures deduced from peaks in chiral
susceptibilities \protect\cite{p4_eos} on lattices of same size.
The right hand side shows  
$\chi_s/T^2$ calculated with p4fat3 and asqtad actions on lattices with
temporal extent $N_\tau =4$, $6$ \protect\cite{Schmidt} and $8$. All results 
are from calculations
with a physical strange quark mass and almost physical, degenerate light
quark masses, $m_l=0.1 m_s$. Data for $N_\tau =8$ are preliminary data of
the hotQCD collaboration \protect\cite{lat07}.
}
\label{fig:sus}
\end{figure}

At small values of the quark chemical potential corrections to the
equation of state can be evaluated systematically in terms of a Taylor
expansion \cite{Taylor} in light ($\mu_l=\mu_u=\mu_d$) and strange 
($\mu_s$) quark chemical potentials.
The expansion coefficients are evaluated at vanishing values for the
chemical potential and are related to thermal fluctuations of light quark and
strangeness numbers in a thermal medium at $\mu_{l,s}=0$.
The leading order, quadratic corrections are given in
terms of the diagonal quark number susceptibilities $\chi_l$ and $\chi_s$
as well as an off-diagonal susceptibility $\chi_{ls}$,
\begin{equation}
\frac{p(T,\mu_l,\mu_s) }{T^4} = \frac{p(T,0,0) }{T^4} 
+\frac{1}{2} \frac{\chi_l}{T^2} \left( \frac{\mu_l}{T} \right)^2
+\frac{1}{2} \frac{\chi_s}{T^2} \left( \frac{\mu_s}{T} \right)^2
+ \frac{\chi_{ls}}{T^2} \frac{\mu_{l}}{T} \frac{\mu_s}{T} + {\cal O}(\mu^4) 
\label{press_mu}
\end{equation}
with
\begin{equation}
\frac{\chi_{q}}{T^2} = \frac{1}{VT^3} 
\frac{\partial^2\ln Z}{\partial(\mu_{q}/T)^2} \; ,\; q=l,s\;\;\; ,\;\;\;
\frac{\chi_{ls}}{T^2} = \frac{1}{VT^3} 
\frac{\partial^2\ln Z}{\partial(\mu_{l}/T)\partial(\mu_{s}/T)} \; \;.
\label{chi_qs}
\end{equation}
In the infinite temperature limit these susceptibilities will approach in
the continuum limit the value for massless two and one flavor quark number 
susceptibilities of an ideal quark gas, respectively,
{\it i.e.} $\lim_{T\rightarrow\infty} \chi_l/T^2 = 2$,
$\lim_{T\rightarrow\infty} \chi_s/T^2 = 1$ and
$\lim_{T\rightarrow\infty} \chi_{ls}/T^2 = 0$.
It is obvious from Fig.~\ref{fig:pfree} that Taylor expansion coefficients
of the pressure, eg. the quark number susceptibilities, suffer from
similar cut-off effects as the pressure evaluated at $\mu_{l,s}=0$ and,
on the other hand, profit also in the same way from a systematic 
${\cal O}(a^2)$ improvement of the action.

Preliminary results for light and strange quark number susceptibilities,
$\chi_{s,l}$, 
obtained with the p4fat3 action on lattices with temporal extent $N_\tau=4$ 
and $6$ \cite{Schmidt} and the asqtad action \cite{milc_Tc} are
shown in Fig.~\ref{fig:sus}. The quark number susceptibilities change 
rapidly in the transition region. Like bulk thermodynamic observables, 
they are sensitive to 
deconfinement and reflect the rapid change from heavy hadronic degrees of 
freedom to light partonic degrees of freedom. In  Fig.~\ref{fig:sus}(right) 
we show in addition preliminary results for the strange quark number
susceptibility obtained by the hotQCD collaboration from calculations 
on a lattice of temporal extent $N_\tau =8$ with the asqtad and p4fat3 
actions \cite{lat07}. These results indicate a 
shift of the transition region to somewhat smaller temperatures, but 
otherwise follow the same pattern seen already in the calculations on 
the coarser $N_\tau =4$, $6$ lattices. We will discuss the deconfining
features as well as the cut-off dependence seen in calculations of 
quark number susceptibilities in more detail in part II of this review.
Here we only point out that $\chi_{l,s}$ have a temperature dependence
very similar to that shown in Fig.~\ref{fig:eps} for the entropy density,
which also is shared by energy density and pressure. For chemical potentials
$\mu_l/T\le 1$ they give the dominant finite-$\mu$ contribution to these
thermodynamic observables, which stays below 10\% for $\mu_l/T\lsim 0.7$.
This clearly is different for higher order derivatives of the partition
function, that are sensitive to thermal fluctuations at non-zero baryon 
number density.

Quark number susceptibilities play an important role in the analysis
of the QCD phase diagram at non-zero baryon chemical potential. 
If there exists a second order phase transition point at non-zero baryon
chemical potential, the baryon (or light quark) number susceptibility
will diverge at this point. It thus is important to get control over
the dependence of $\chi_{l,s}$ on $\mu_{l}$. The leading order, 
quadratic corrections to $\chi_l$ and $\chi_s$ are given in terms of 
fourth order expansion coefficients of the pressure \cite{c6}; for  
vanishing $\mu_s$ one finds, 
\begin{eqnarray}
\frac{\chi_l(\mu_l)}{T^2} = \frac{\chi_l(0)}{T^2} + 
12 c_{40} \left( \frac{\mu_l}{T} \right)^2 
+{\cal O}(\mu_l^4) \;\; &\; ,\;&\;\; 
\frac{\chi_s(\mu_l)}{T^2} = \frac{\chi_s(0)}{T^2} +  
2 c_{22} \left( \frac{\mu_l}{T}\right)^2  
+{\cal O}(\mu_l^4) \; ,
\label{suscept_mu}
\end{eqnarray}
with expansion coefficients 
$c_{nm} = (VT^3)^{-1}\left[ \left(\partial \ln{\rm Z}(T,\mu_l,\mu_s)/\partial 
(\mu_l/T)^n\partial (\mu_s/T)^m \right)/n!/m! \right]_{\mu=0} $.

The expansion coefficients, $c_{40}$ and $c_{22}$, are positive and
have a maximum at the temperature of the transition at vanishing $\mu_{l,s}$.
For non-zero light quark chemical potential the susceptibilities thus 
increase and start developing a peak that becomes more pronounced
with increasing $\mu_l$. 
\begin{figure}[t]
\begin{center}
\hspace*{-0.4cm}\epsfig{file=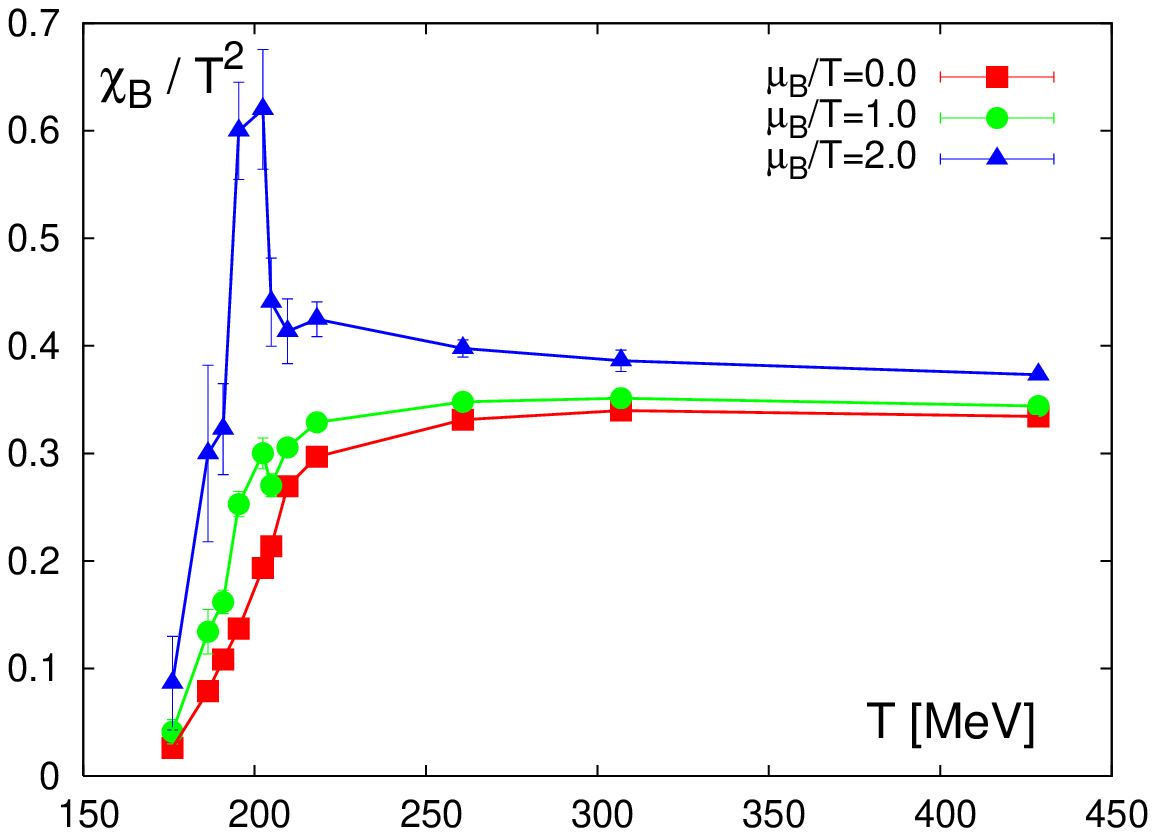, width=7.9cm}
\hspace*{-0.4cm}\epsfig{file=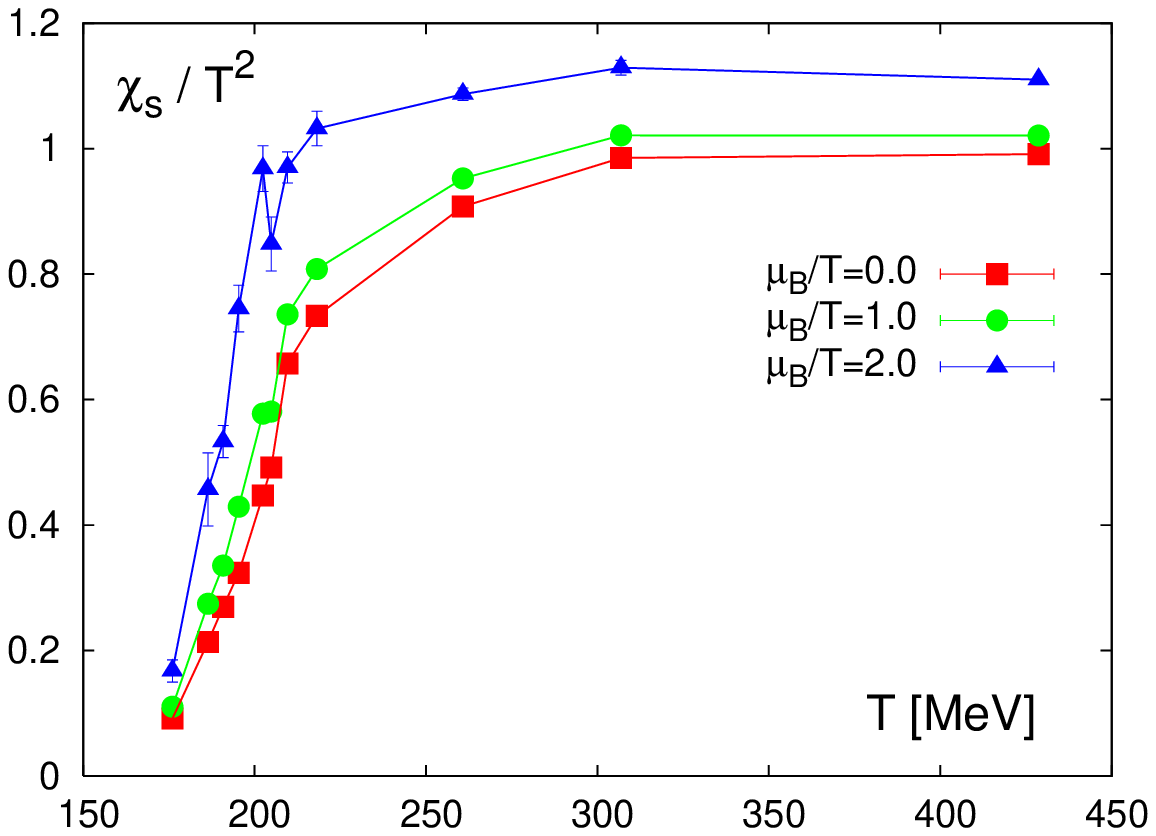,width=7.9cm}
\end{center}
\caption{Fluctuations of baryon number (left) and strangeness (right)
at non-zero baryon chemical potential calculated in leading order
Taylor expansion on lattices with
temporal extent $N_\tau=4$ in QCD with a physical strange quark mass
and almost physical light quark masses, $m_l =0.1 m_s$.
}
\label{fig:fluctuation}
\end{figure}
This is obvious from Fig.~\ref{fig:fluctuation} where we show
baryon number and strangeness susceptibilities rather than
light and strange quark number susceptibilities. The former 
are obtained in analogy to the latter after replacing in the
QCD partition function the $(u,d,s)$ chemical potentials by
appropriate linear combinations of $(B,Q,S)$ chemical potentials
\cite{Schmidt,Gavai,Heller}. The leading order Taylor expansion
for the baryon number susceptibility 
evaluated here as function of the baryon chemical potential, 
$\mu_B = 2\mu_d+\mu_u\equiv3\mu_l$,
for vanishing strange quark chemical potential, $\mu_S=\mu_d-\mu_s$, and 
also for
vanishing charge chemical potential, $\mu_Q=\mu_u-\mu_d$, is then given 
by \footnote{Other choices for the chemical potentials may be more
suitable to resemble situations encountered, for instance, in heavy ion
collision. Charge and strangeness chemical potentials may
be adjusted to reproduce conditions met in a particular event-by-event
analysis of fluctuations \cite{Begun}.},
\begin{equation}
\chi_B(\mu_B) = \frac{1}{9}\left( \chi_l(\mu_B) +\chi_s(\mu_B) +
2 \chi_{ls}(\mu_B) \right) \; . 
\label{chiB}
\end{equation}
We note that strangeness fluctuations do react to an increase in net
baryon number ($\mu_B>0$). This also holds true when the strange quark
chemical potential is not chosen to be zero but is tuned such to insure
that overall strangeness vanishes \cite{Schmidt,Heller}. The 
preliminary results shown in Fig.~\ref{fig:fluctuation} so far have only
been obtained on lattices with temporal extent $N_\tau=4$. 
Of course, one has to check the cut-off dependence also in this case through
calculations on larger lattices \cite{Schmidt}. Moreover, truncation effects 
in the Taylor expansion need to be analyzed by going beyond
leading order in the expansion.
In fact, higher order corrections are important! Only in the next order does
the position of the peak become sensitive to the chemical potential and 
will get shifted towards lower temperatures with in creasing $\mu_l$.

\section{Conclusions}

Using ${\cal O}(a^2)$ improved staggered fermion actions with an almost
physical quark mass spectrum much progress has been made in calculating 
the equation of state of QCD and extracting phenomenologically important
quantities like the velocity of sound and fluctuations of hadronic charges.
The current studies, performed mainly on lattices with temporal extent
$N_\tau =4$ and $6$ show some cut-off dependence. Calculations performed
at high temperature for the trace anomaly on lattices with temporal extent
$N_\tau=8$ suggest that these effects are small. A more detailed analysis,
in particular at smaller temperatures in the hadronic phase, however, is 
still needed before firm conclusions on the EoS in the continuum limit
can be drawn.  
Preliminary results obtained for quark number susceptibilities show that
the transition region shifts to smaller temperatures with increasing $N_\tau$.
This, of course, in accordance with the shift seen in systematic studies
of the transition temperature performed with the p4fat3 \cite{p4_Tc} and
asqtad \cite{milc_Tc}. actions. To quantify the effect for the equation of
state will require a more detailed study on lattices with temporal extent
$N_\tau =8$.

\end{document}